\documentclass[sigconf]{acmart}

\AtBeginDocument{%
  \providecommand\BibTeX{{%
    \normalfont B\kern-0.5em{\scshape i\kern-0.25em b}\kern-0.8em\TeX}}}

\renewcommand\footnotetextcopyrightpermission[1]{} 
\usepackage{multirow}
\usepackage{enumitem}

\begin{document}

\title{G-Issue: Analyzing Lifetime and Evolution of Issue-related Artifacts from Open Source Repositories}

\author{Sayed Mohsin Reza}
\orcid{0000-0003-3379-6319}
\affiliation{%
  \institution{University of Texas at El Paso}
  \city{El Paso}
  \state{Texas}
  \postcode{79902}
  \country{USA}
}
\email{sreza3@miners.utep.edu}

\author{Saif Uddin Mahmud}
\affiliation{%
  \institution{University of Texas at El Paso}
  \city{El Paso}
  \state{Texas}
  \postcode{79902}
  \country{USA}
}
\email{smahmud4@miners.utep.edu}

\author{Omar Badreddin}
\affiliation{%
  \institution{University of Texas at El Paso}
  \city{El Paso}
  \state{Texas}
  \postcode{79902}
  \country{USA}
  }
\email{obbadreddin@utep.edu}


\renewcommand{\shortauthors}{Reza et al.}

\begin{abstract}
Software developers or contributors report issues related to bugs, errors, and missing documentation during community-based software development. These issues are treated as feedback and are crucial to enhancing software new features, documentation, and quality. If software issues are not being addressed with a correct developer, software quality degrades and is unable to use in the end. Hence, it is essential to analyze the software issue-related artifacts to understand the behavior of the software. This paper investigates the performance of the proposed issue-related artifacts mining tool G-Issue with other state-of-the-art tools. We also investigate issue lifetime and evolution of issues over time among well-known and maintained repositories. The results show that  G-Issue is faster in mining issue-related artifacts but takes more memory than general Python API during mining issue mining. The results depict that we can prioritize issues based on issue lifetime and evolution. Such results may provide a new horizon about issues that can help in issue management, developer assignment, and quality management.

G-Issue URL: \url{https://www.smreza.com/projects/modelmine/issues.php}

\end{abstract}




\maketitle

\section{Introduction}
Software development becomes distributed nowadays, and developers from anywhere can contribute towards the software development \cite{begel2013social}. Towards this development, some software manages technical artifacts like commits, issues, and milestones which enables a social community that attracts many developers to work on and deliver projects within timeline \cite{teixeira2019managing,bertram2010communication}. BitBucket  \cite{fisher2013utilizing}, GitHub\cite{blischak2016quick}, GitLab \cite{rios2019methodology} is the leader in distributed version control and source code management (SCM), which combines the ability to develop, secure, and operate software in a single application. 

Source code management software is growing in features that allow faster development through bug identification, error reporting, or other issues. One of the features is an issue tracking system, often used to get user feedback related to proposed features, bugs, errors, and problems. Also, the service allows the developers to assign an issue to a developer \cite{liao2018exploring} and automatic labeling issues to prioritize it better \cite{kallis2019ticket}. In summary, this tracking system enhances the code quality and increases the software lifetime. 

Software maintenance is a costly and largely unpredictable human-intensive activity in the software development life cycle. High maintenance efforts and expertise often eclipse the cost and sometimes become the reason for unsustainable software \cite{hatton2017long}. Moreover, if issues are not well managed during this maintenance, the software becomes smelly and may introduce bugs, and obsolete in the long run \cite{rodriguez2016bugtracking}. To solve such issues, developers worldwide may provide feedback on an issue and can contribute to fixing that. Therefore, source code management with issue tracking can provide collaborative pathways to manage software,  reduce software failures and improve software quality. 

Very few research efforts have been conducted on mining \cite{jurado2015sentiment,duenas2018perceval,sun2015msr4sm}, analyzing \cite{de2016systematic} and visualizing  \cite{fiechter2021visualizing} issues in open source communities. These efforts include issue title prediction \cite{zhang2022itiger}, automatic issue labeling \cite{wang2021well,wang2022personalizing} and sentiment analysis of issues \cite{guzman2014sentiment,ding2018entity}. However, there is a missing effort on mining issues faster, analyzing issue timelines, and evolving issues over time. 

In this paper, we investigate the performance of issue mining of an in-house developed tool, called G-Issue, and compare performance with other state-of-the-art tools \cite{spadini2018pydriller,gousios2012ghtorrent} in terms of execution time and memory usage. Moreover, we investigate the average issue lifetime in popular open source repositories and analyze the evolution of issues over time among repositories to see the behavior of each repository.

This study is structured as follows: Section \ref{related_work} discusses the related research works on issue tracking and its management;  Section \ref{study_design} discusses the study design with research questions. Section \ref{results} shows results against each research questions and discuss elaborately and finally we conclude in Section \ref{conclusion}.

\section{Related Work}
\label{related_work}

Software development through source code management and its associated artifacts are available on an open-source platform. Several studies have been conducted research on such artifacts from different perspective such as sentiment analysis \cite{jurado2015sentiment}, label prediction\cite{kikas2016using}, issue management \cite{bissyande2013got} \& mining \cite{zhang2022itiger}.

\subsection{Issue-related Artifact Mining}
Software artifact mining has improved software quality, bug identification, and network analysis. Several studies have uncovered interesting and actionable artifacts from software data. Several mining tools have emerged to enable such research, and discovery \cite{spadini2018pydriller,reza2020modelmine, romano2021g}. For example, PyDriller, a python framework for mining software repositories, can extract recent information from open source repositories such as commits, developer information, modifications, differences, and source codes \cite{spadini2018pydriller}. However, the tool has no feature to extract issues. MetricMiner is another
application suitable for mining software repositories for metrics calculation, data extraction, and statistical inference \cite{sokol2013metricminer}. These tools focus on extracting data primarily from either code or commit history, with limited support for mining issue-related artifacts. 

\subsection{Issue-related Artifact Analysis}
Issue-related artifact analysis has gained popularity last few years. Several studies have researched on issue lifetime \cite{kikas2016using,rees2017better}, how long it will take to close an issue, and empirical studies on the life expectancy of issues based on labels. Kikas et al. conducted research on 4000 repositories to find temporal dynamics of issues in GitHub \cite{kikas2016using}. The study found that the projects with a shorter observation time tend to have higher volumes of open issues. In addition, Kikas proposed a prediction model trained from static, dynamic, and contextual features to predict the lifetime of an issue. The results showed that the average issue lifetime for community-created issues is 39 days, but the team-created issues are 5.9 days.

\subsection{Issue-related Artifact Visualization}
Visualization techniques have been popular in textual \& network data. As issue-related artifact has textual(issue title, body) and network (issue assignee, label) data, several research have been conducted on issue-related artifact visualization \cite{liao2018exploring,bissyande2013got}. Liao et al. applied the visualization technique on issue-related behavior by analyzing seven projects with 98,074 issues in total \cite{liao2018exploring} and proposed an SRF to measure the importance of user behaviors. The results found that issue-related user behaviors are critical, and not all issues that are assigned labels could be closed rapidly. Another empirical research by Bissyand et al. investigated the adoption of an issue tracker based on the projects, developers, and type of issues \cite{bissyande2013got}. The results found that small-sized team projects are less likely to have issues and visualized the top 10 labels in GitHub, where bug and feature requests are at the top with 18.36\% and 10.29\%, respectively.

\subsection{Issue-related Other Research}
There are other research on issues conducted in aspects of sentiment \cite{ding2018entity,jurado2015sentiment,yang2017sentiments}, bug and issue-tracker analysis \cite{rodriguez2016bugtracking}. Jurado et al. conducted a study on 10,829 issues from 9 well-known \& open source repositories to do sentiment analysis \cite{jurado2015sentiment}. The study proposed a new technique to identify the underlying sentiments in the text found in issues and their comments. Results showed that issues' titles and text leave underlying sentiments, which can be used to analyze the development process. Another research on issues-related artifacts is automatic labeling of GitHub Issues \cite{golzadeh2021ground,kallis2019ticket,bharadwaj2022github}. Kallis et al. introduced a tool called Ticket Tagger, which is developed using Node.js and de-facto server-side JavaScript and uses machine learning approaches on issue artifacts to label an issue automatically  \cite{kallis2019ticket}.

In summary, there is insufficient research on issue-related artifact mining, analysis, and visualization through a web application. Hence, this scope study is unique in terms of faster mining without any hassle of processing, analyzing, and visualizing the artifacts.

\section{Study Design}
\label{study_design}
The study aims to analyze issue-related artifacts from open source repositories with the purpose of mining, pre-processing, and visualizing the issues which can be effectively used in practice. The perspective is of both researchers and practitioners who are interested in analyzing the issues in terms of issue expectancy and evolution of issues. Specifically, we aim to address the following research question:

\begin{figure*}[t]
\centering
 \includegraphics[width=0.9\linewidth]{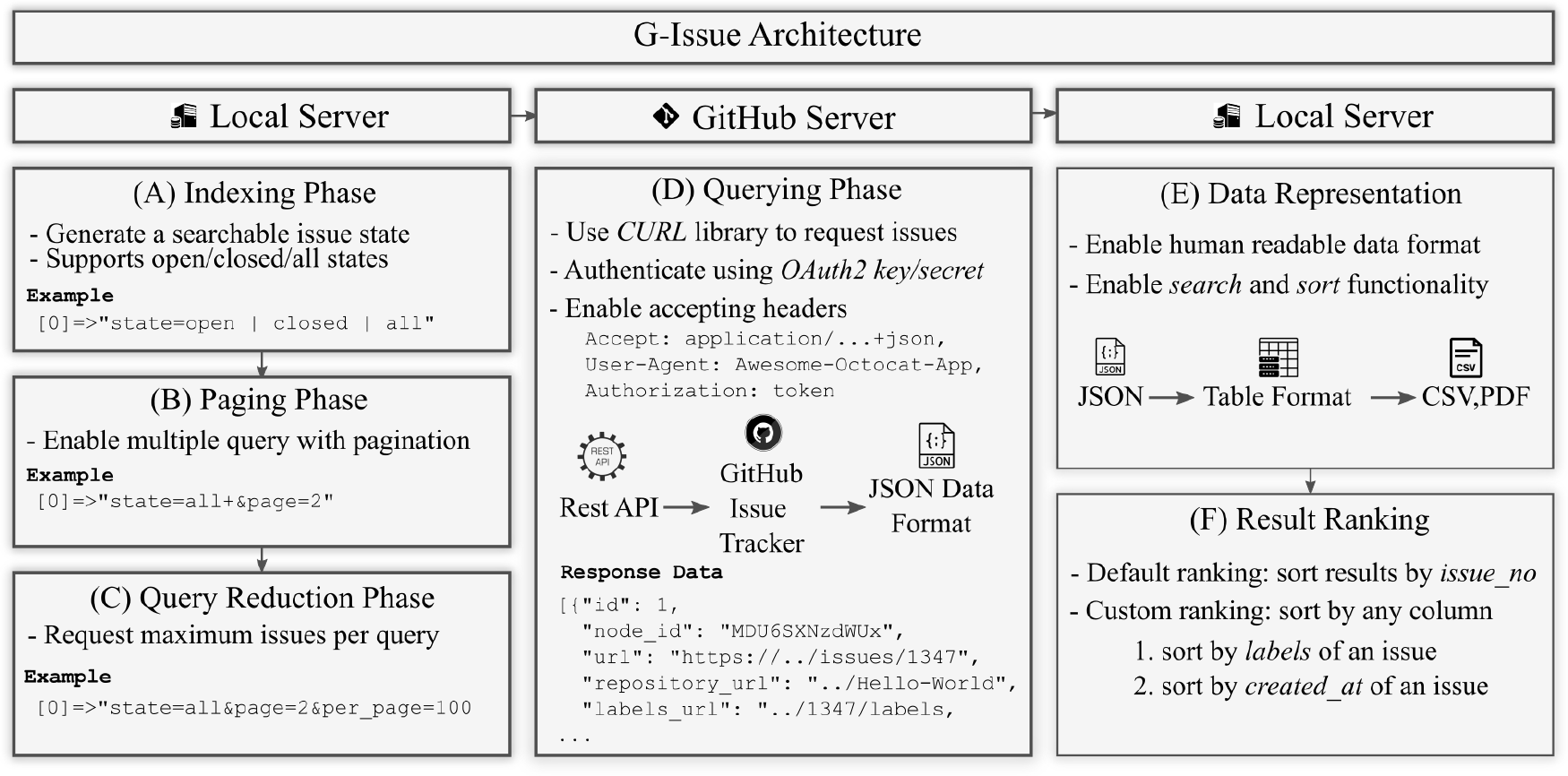}
 \caption{Architecture of G-Issue Tool }
\label{fig:Architecture}
\end{figure*}

\subsection{Research Questions}
\label{research_questions}

This section discusses the research questions we used and how we plan to answer these research questions. We are motivated to find the answer to the following research questions:

\textbf{RQ1.} What is the performance of the G-Issue tool compared to the state-of-the-art tools in mining issue-related artifacts?

The RQ focuses on the performance evaluation of G-Issue and is motivated by the fact that issue-related artifacts are crucial in repositories compared to code itself and tend to be significantly larger in terms of text size and issue comments. This often translates to complexity in identifying and extracting issue-related artifacts. For reference, we compare ModelMine with state of the art tools Python API \cite{jurado2015sentiment},  GHTorrent \cite{gousios2012ghtorrent,gousios2014lean}, PyDriller \cite{spadini2018pydriller}, G-Repo \cite{romano2021g} for mining issues from GitHub. To answer this research question, we choose three individual tasks that are common for the majority of mining research with available support in mining tools. The tasks are as follows:

\begin{enumerate}
 \item \textbf{Task 1 (Size related):} Retrieve the list of 1000 issues that include at least one open state issue, and the total number of issues is more than 1000.
 \item \textbf{Task 2 (Time-related):} Retrieve the list of 1000 issues that include at least one open state issue and created before January 2019.
 \item \textbf{Task 3 (State related):} Retrieve the list of 1000 issues now in a closed state.
\end{enumerate}

These tasks are implemented using the following frameworks/tools: (1) G-Issue, (2) Python API, (3) GHTorrent, (4) PyDriller, and (5) G-Repo. To compare the tools, we use two performance metrics: (1) Execution Time and (2) Max Memory (MM). Such performance metrics are used in evaluating different software artifacts mining tools \cite{dyer2013boa,spadini2018pydriller,reza2020modelmine}. The evaluation checks how fast and how much memory the tool takes to mine issue-related artifacts.

\textbf{RQ2.}  What is the average issue lifetime among different repositories?

This RQ describes the analysis of the time it takes to solve an issue for each repository in our dataset on average. After collecting issues using G-Issue, we will find closed state issues, its created time, and when it is closed. We reveal the average issue lifetime among different repositories based on those data.

\textbf{RQ3.}  What is the evolution of issues over time among repositories?

This RQ shows the evolution of open and closed state issues among repositories. To prepare the results, we need to extract yearly issues and their state from all the issues. We also plan to show Kernel Density Estimation (KDE) as a part of the probability density function on our ongoing issue creating time variables.

With these research questions, we aim to provide a more profound knowledge of the capabilities of G-Issue in mining and analysis of issue-related artifacts. The following subsections report the architecture of G-Issue and the steps that we conducted to collect the dataset.

\subsection{G-Issue Architecture}
\label{architecture}
In this section, we discuss the architecture of the issue mining tool G-Issue that we built in-house lab setup and hosted on the online platform. The tool adopts several approaches (indexing, paging, query reduction, querying, data representation, and results ranking) to mine issue-related artifacts of repositories from open source repositories. The overall architecture of the G-Issue tool
is visualized in Figure \ref{fig:Architecture}.

In G-Issue, we provide a user interface with the mining capability to request GitHub for issue-related artifacts and process that data. This service is under the parent tool called ModelMine \cite{reza2020modelmine}. This tool provides a simple, extensible user interface to mine issue-related artifacts of repositories. It has a different way of searching to ensure the possibility of different mining types of datasets for MSR research.

\begin{figure}[!htb]
\centering
 \includegraphics[width=0.9\linewidth]{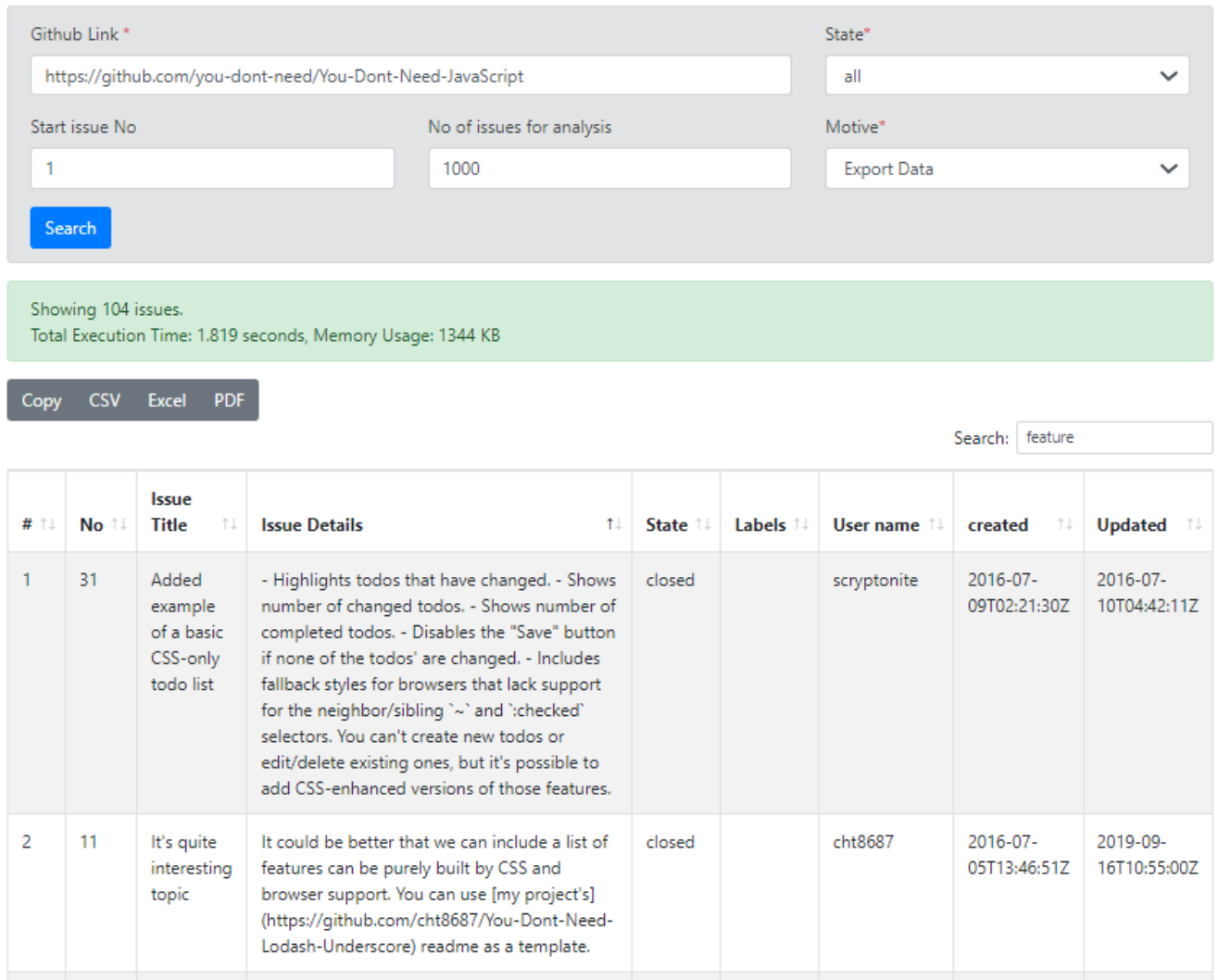}
 \caption{Search \& result screenshot of G-Issue Tool }
\label{fig:screenshot_g-issue}
\end{figure}

Software issues have multiple types of artifacts, including state, milestone, assignee, and G-Issue, allowing researchers to investigate specific state-based issue searches. This feature allows researchers to analyze the different states of the issues in repositories and the behavior of software code issues of different projects. The user interface of the G-Issue tool is visualized in Figure \ref{fig:screenshot_g-issue}.

\subsection{Data Collection}
\begin{table*}[t]
\caption{Selected repositories with metadata information}
\label{tab:repositories}
\resizebox{\textwidth}{!}{%
\begin{tabular}{llrrrrcp{0.1\textwidth}p{0.1\textwidth}p{0.1\textwidth}}
\hline
\textbf{Serial} & \textbf{Repository name} & \textbf{Commits} & \textbf{Contr.} & \textbf{Stars} & \textbf{Forks}& \textbf{Time Selection} & \textbf{No. Open Issues} & \textbf{No. Closed Issues} & \textbf{Total Issues} \\ \hline
1 & Spring framework  & 22,208 & 531 & 41,400 & 28,800 & 2004-05 to 2022-08 & 1,391 & 24,593 & 25,985 \\
2 & Junit-5 & 6,621 & 161 & 4,400 & 991 & 2015-01 to 2022-08 & 135 & 2,828 & 2,963 \\
3 & Apache kafka & 8,590 & 762 & 18,000 & 9,600 & 2012-08 to 2022-08 & 1,002 & 11,477 & 12,479  \\
4 & Apache lucene-solr & 34,789 & 232 & 4,100 & 2,700 & 2016-01 to 2022-08 & 255 & 2,411 & 2,666  \\
5 & Dropwizard & 5,702 & 361 & 7,900 & 3,300 & 2011-03 to 2022-08 & 26 & 5,518 & 5,544  \\
6 & Checkstyle & 9,922 & 254 & 5,800 & 7,700 & 2013-09 to 2022-08 & 697 & 11,287 & 11,984  \\
7 & Hadoop & 24,612 & 339 & 11,300 & 7,000 & 2014-09 to 2022-08 & 681 & 3,681 & 4,362  \\
8 & Selenium & 26,532 & 558 & 19,800 & 6,200 & 2013-01 to 2022-08 & 117 & 10,597 & 10,715  \\
9 & Skywalking & 6,242 & 315 & 16,100 & 4,700 & 2015-11 to 2022-08 & 62 & 8,525 & 8,587  \\
10 & Signal android & 7,015 & 223 & 19,800 & 4,700 & 2011-12 to 2022-08 & 242 & 10,049 & 10,291  \\ \hline

\end{tabular}
}
\end{table*}

One of the challenges in software research is identifying code repositories that have been actively maintained for an extended period. We identify some characteristics that may give us actively maintained repositories to search such code repositories. The characteristics are as follows: a repository with a minimum of 5000 commits, at least 100 active contributors, a minimum of 3000 stars, and 500 forks. We use the ModelMine tool \cite{reza2020modelmine} which is capable of retrieving repositories with the mentioned criteria. A high number of stars and forks imply the popularity of the repositories, and a high number of commits imply maintenance throughout the software development life cycle. We choose the top ten repositories from the results provided by the ModelMine tool. Overall, the selected repositories have code changes in commits that will help us to extract the different source code metrics to reduce threats to the generalizability of this study. In this study, we have mined repositories and created a dataset composed of ten open source repositories. Then we use the G-Issue tool to mine issue-related artifacts. The whole dataset is now published in GitHub \url{https://github.com/sayedmohsinreza/CSIQ} and available online \cite{reza2022csiq}. The detailed summary of the ten open source repositories and issues in each repository are reported in Table \ref{tab:repositories}.

\subsection{Terminology}
The software issues have some particular terminology we need to discuss to understand the results. Occasionally, issue-related artifacts include reporting bugs, requesting new features, refactoring code, and enhancement ideas. Also, the artifacts are typically created by anyone with title 
\& details and consist of the person's information, created time, and labels associated with the issues. If the issue is closed or modified, that record is also documented in the specific issue.

Here are the details of some terminologies used in this study.
\begin{itemize}
    \item \textbf{Issue lifetime -} Time from the first opening of the issue to the first closing of the issue.
    \item \textbf{Opened issue -} Newly created issue. Each issue is opened only once during its lifetime.
    \item \textbf{Closed issue -} issue that is marked closed in the issue tracking system. In practice, an issue might be reopened and closed again, but here we use only the last closing event.
\end{itemize}

\section{Results \& Discussion}
\label{results}
In this section, we report the results and analysis of the research questions mentioned in Section \ref{research_questions}.

\subsection{Performance Evaluation}
This section discusses the results of the performance of G-Issues compared to other state-of-art-tools. The performance evaluation results among the tools are visualized in Table \ref{tab:performance-results}. Such a result provides an idea of which tool performs better during mining issue-related artifacts and how much fast and memory the tool takes to mine selected repositories. All these results are produced with the setup to mine 1000 issues from repositories.

\begin{table}[!htb]
\centering
\caption{Performance evaluation results}
\label{tab:performance-results}
\resizebox{\columnwidth}{!}{
\begin{tabular}{lllllll}
\hline
\textbf{Tasks} & \textbf{Metrics} & \textbf{G-Issue} & \textbf{Python API} & \textbf{GHTorrent} & \textbf{PyDriller} & \textbf{G-Repo} \\ \hline
Task $1^{*}$  & $ET^{**}$ & \textbf{12.1s} & 18.2s & 46.2s & Not  & Not  \\
(Size) & $MM^{***}$ & 18223KB & \textbf{10211KB} & 67033KB & supported   & supported  \\ \hline
Task 2 & ET &  \textbf{30.22s} & 41.7s & 88.3s & Not  & Not \\
(Time) & MM & 20340KB & \textbf{16547KB} & 74031KB & supported   & supported  \\ \hline
Task 3  & ET & \textbf{11.8s} & 15.5s & 102.3 & Not & Not \\
(Issue-related) & MM & 19967KB & \textbf{14566KB} & 63654KB & supported   & supported   \\ \hline
\multicolumn{7}{l}{* Task details are listed in Section \ref{research_questions}} \\
\multicolumn{7}{l}{** ET - Execution Time} \\
\multicolumn{7}{l}{*** MM - Max Memory} \\
\end{tabular}
}
\end{table}

Table \ref{tab:performance-results} shows that in each task, G-Issue mines a list of 1000 issues with the lowest execution time while GHTorrent mines with the highest execution time. Python API has the lowest memory utilization during mining, and GHTorrent has the highest utilization. Among state-of-the-art tools, PyDriller and G-Repo have focused on mining software repositories and have no feature to mine issue-related artifacts.

\subsection{Analysis of Issue Lifetime}
In this section, the results of issue lifetime among repositories are discussed and portrayed in Table \ref{tab:average_days}. The table shows the average days it takes to solve an issue among repositories. Here "Average days to solve" means how many days it takes to close the issue by developers since the issue creation date. 

\begin{table}[!htb]
\caption{Statistics on days it takes to solve an issue}
\label{tab:average_days}
\resizebox{\linewidth}{!}{%
\begin{tabular}{lccc}
\hline
\textbf{Project Name} & \textbf{Mean (days)} & \textbf{Minimum (days)} & \textbf{Maximum (days)} \\ \hline
1. spring-framework & 1220 & 0 & 5491 \\
8. selenium & 551 & 0 & 2574 \\
10. signal-android & 215 & 0 & 3010 \\
2. junit-5 & 162 & 0 & 2144 \\
3. apache-kafka & 104 & 0 & 2467 \\
5. dropwizard & 101 & 0 & 3221 \\
7. hadoop & 98 & 0 & 2158 \\
4. apache-lucene-solr & 91 & 0 & 2030 \\
6. checkstyle & 57 & 0 & 2496 \\
9. skywalking & 37 & 0 & 1840 \\ \hline
\end{tabular}%
}
\end{table}

From Table \ref{tab:average_days} results, we can see that \emph{Spring Framework} project has the highest on average of 1220 days to solve an issue where \emph{skywalking} developers use only 37 days. \emph{spring-framework} commit count is less than \emph{haddop} project but average issue lifetime in \emph{haddop} is twelve time less than \emph{spring-framework}. For each repository, the minimum issue lifetime day is zero, which implies that within the issue created date, developers solve the issue and close that.

\begin{figure}[!htb]
\centering
 \includegraphics[width=\linewidth]{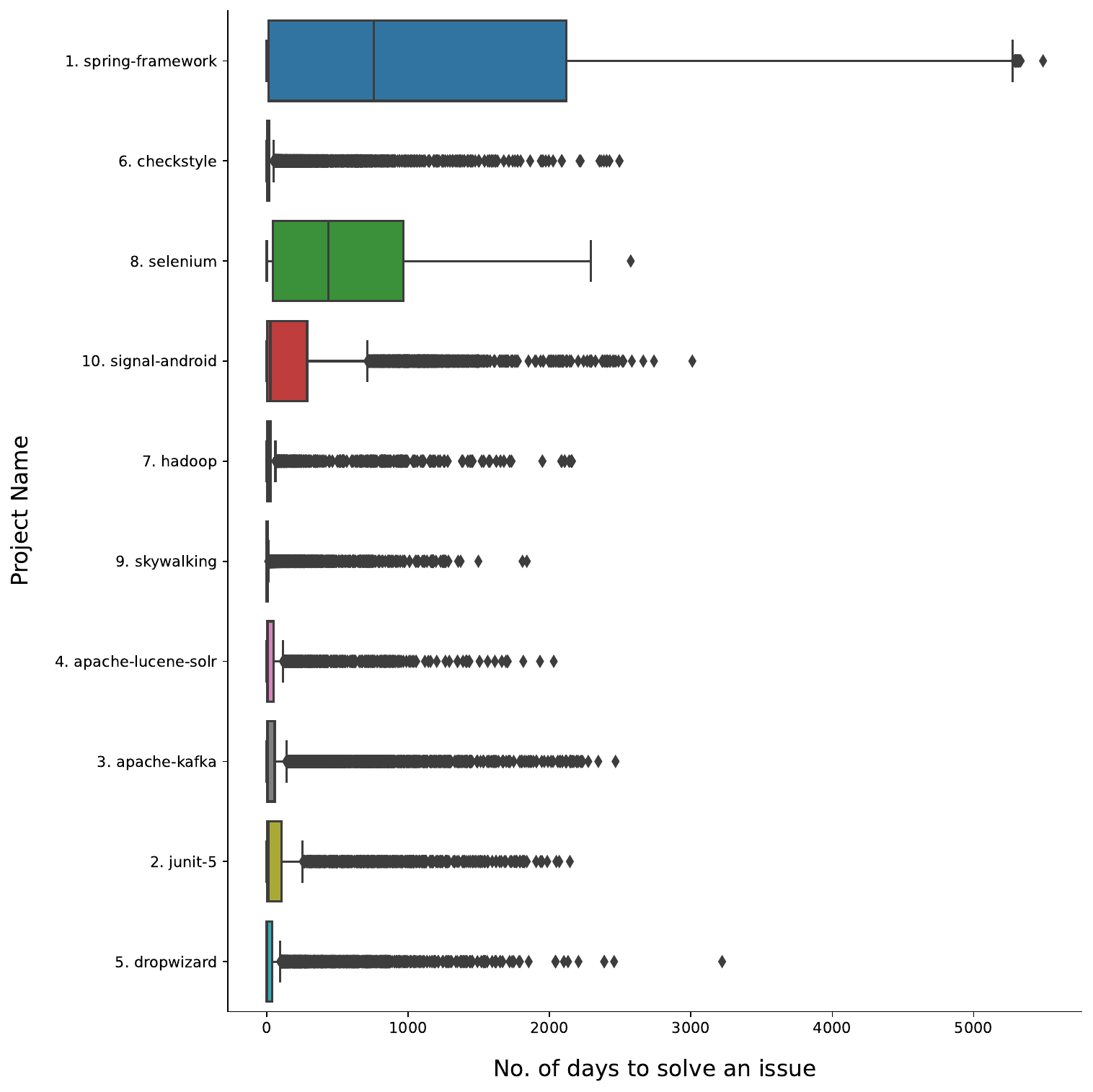}
 \caption{Box plot of days it takes to solve issues among repositories}
\label{fig:average_issule_lfetime}
\end{figure}

However, Figure \ref{fig:average_issule_lfetime} visualizes the boxplot of issue lifetime among repositories. From the figure, it is noticeable that \emph{spring-framework} and \emph{selenium} has the highest mean of days to solve an issue. Among all repositories, one issue from \emph{spring-framework} has taken more than 5000 days / 13 years to solve. Here, we need to keep in mind that some issues are closed and reopened later on to receive more feedback on that issue.

\subsection{Evolution of Issues}
\begin{figure}[t]
\centering
 \includegraphics[width=1.1\linewidth]{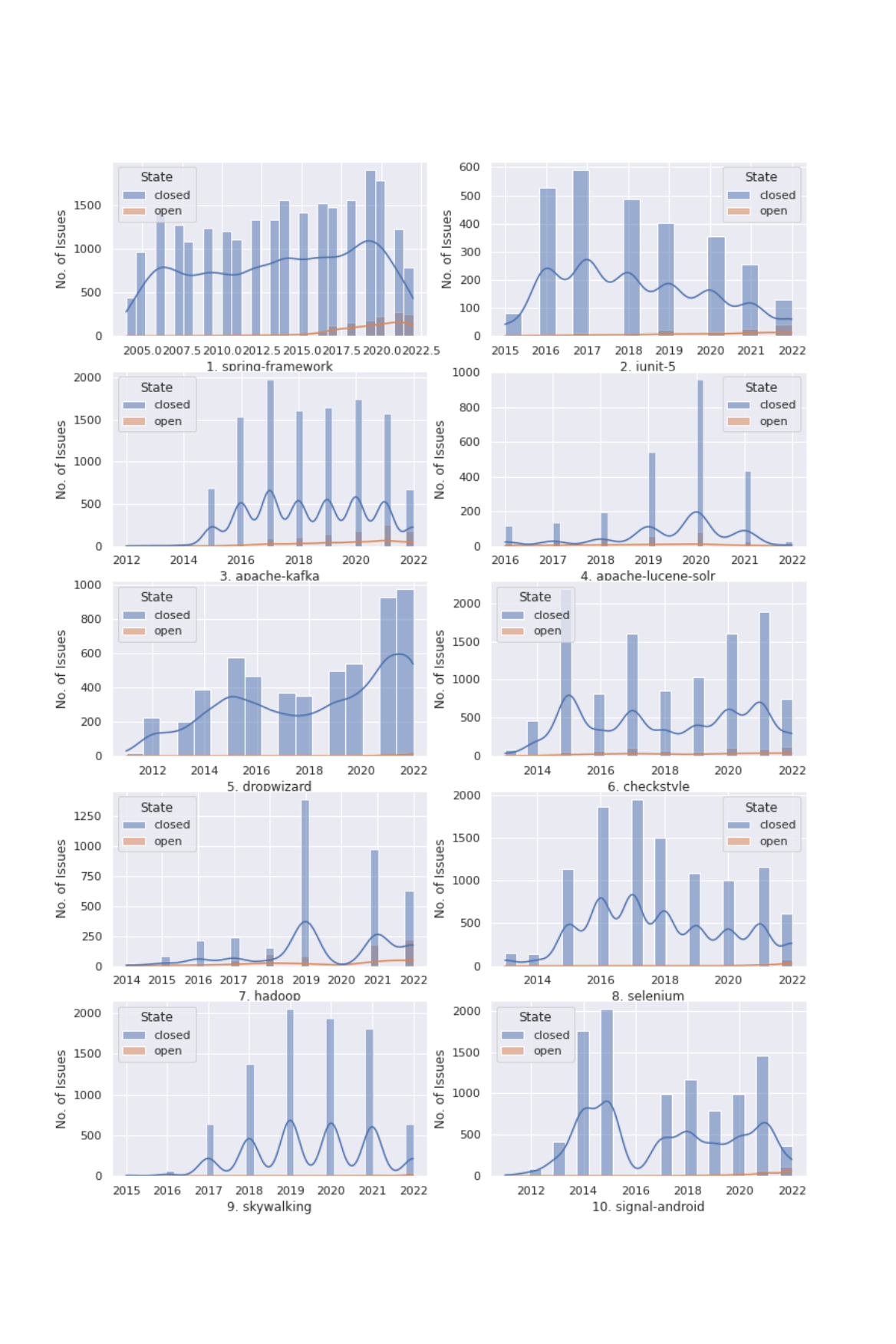}
 \caption{evolution of issue-related artifacts over time among repositories}
\label{fig:evolution}
\end{figure}

In this section, we discuss the evolution of issues among repositories. The results of the evolution of issues are visualized in Figure \ref{fig:evolution} showing a histogram of issue count per year in terms of open or closed state among repositories. 

In every case, the graph implies that new issues are increasing in number during the software evolution. This number increases and becomes higher when the close-state issue rate declines. \emph{spring-framework}, \emph{junit-5}, \emph{checkstyle} and \emph{signal-android} shows a recent decline in the rate of closed-state issues and an upward trend of new issues. The KDE density value represents an increasing number of issues reported by developers or contributors. 

Also, we have seen a pattern of the zigzag move of issues over the years among the repositories. It implies that when new issues are introduced within that year, it tries to be solved and closed the issue. Hence, the continuous maintenance through issue-related artifact analysis prepares software for the subsequent releases with improved software quality and minimized bugs in reporting.

\section{Conclusion}
\label{conclusion}
Software maintenance is crucial during software development. If the maintenance efforts are not made correctly, the software quality degrades over time and is hard to fix at one point. To do software maintenance, developers need feedback in the form of issues. Most source code management software nowadays provides issues to report bugs and share ideas for new features.

In this study, we investigated the process of mining, analyzing, and visualizing issue-related artifacts through a developed tool called G-Issue. The study primarily compares the performance of the G-Issue tool with state-of-the-art tools. Moreover, we investigate the lifetime and evolution of issues in well-known open source projects. 

The results show that the G-Issue tool performs a minimum of 33\% faster than other state-of-the-art tools. However, in memory management, G-Issue is higher than the Python API but lower than other tools. Besides, the results show that highly popular \& forked repositories have more issues; on average, it takes more days to solve an issue. In terms of evolution, if the rate of the closed issue is declining, there is a high chance of introducing new issues. Such results may provide new knowledge about issues-related artifacts and help team leaders with issue assignments for better software development. 

In future research, we plan to analyze the issue text and apply natural language processing to identify issue labels, improving the issue label tracking system.

\bibliographystyle{ACM-Reference-Format}
\bibliography{papers}

@inproceedings{kikas2016using,
  title={Using dynamic and contextual features to predict issue lifetime in github projects},
  author={Kikas, Riivo and Dumas, Marlon and Pfahl, Dietmar},
  booktitle={2016 ieee/acm 13th working conference on mining software repositories (msr)},
  pages={291--302},
  year={2016},
  organization={IEEE}
}

@inproceedings{de2016systematic,
  title={A systematic mapping study on mining software repositories},
  author={de F. Farias, M{\'a}rio Andr{\'e} and Novais, Renato and J{\'u}nior, Methanias Cola{\c{c}}o and da Silva Carvalho, Lu{\'\i}s Paulo and Mendon{\c{c}}a, Manoel and Sp{\'\i}nola, Rodrigo Oliveira},
  booktitle={Proceedings of the 31st Annual ACM Symposium on Applied Computing},
  pages={1472--1479},
  year={2016}
}

@article{zhang2022itiger,
  title={itiger: An automatic issue title generation tool},
  author={Zhang, Ting and Irsan, Ivana Clairine and Thung, Ferdian and Han, DongGyun and Lo, David and Jiang, Lingxiao},
  journal={arXiv preprint arXiv:2206.10811},
  year={2022}
}

@inproceedings{rodriguez2016bugtracking,
  title={Bugtracking: A tool to assist in the identification of bug reports},
  author={Rodr{\'\i}guez-P{\'e}rez, Gema and Gonzalez-Barahona, Jes{\'u}s M and Robles, Gregorio and Dalipaj, Dorealda and Sekitoleko, Nelson},
  booktitle={IFIP International Conference on Open Source Systems},
  pages={192--198},
  year={2016},
  organization={Springer}
}

@article{hatton2017long,
  title={The long-term growth rate of evolving software: Empirical results and implications},
  author={Hatton, Les and Spinellis, Diomidis and van Genuchten, Michiel},
  journal={Journal of Software: Evolution and Process},
  volume={29},
  number={5},
  pages={e1847},
  year={2017},
  publisher={Wiley Online Library}
}

@article{begel2013social,
  title={Social networking meets software development: Perspectives from github, msdn, stack exchange, and topcoder},
  author={Begel, Andrew and Bosch, Jan and Storey, Margaret-Anne},
  journal={IEEE software},
  volume={30},
  number={1},
  pages={52--66},
  year={2013},
  publisher={IEEE}
}

@inproceedings{ding2018entity,
  title={Entity-level sentiment analysis of issue comments},
  author={Ding, Jin and Sun, Hailong and Wang, Xu and Liu, Xudong},
  booktitle={Proceedings of the 3rd International Workshop on Emotion Awareness in Software Engineering},
  pages={7--13},
  year={2018}
}

@inproceedings{guzman2014sentiment,
  title={Sentiment analysis of commit comments in GitHub: an empirical study},
  author={Guzman, Emitza and Az{\'o}car, David and Li, Yang},
  booktitle={Proceedings of the 11th working conference on mining software repositories},
  pages={352--355},
  year={2014}
}

@inproceedings{fiechter2021visualizing,
  title={Visualizing github issues},
  author={Fiechter, Aron and Minelli, Roberto and Nagy, Csaba and Lanza, Michele},
  booktitle={2021 Working Conference on Software Visualization (VISSOFT)},
  pages={155--159},
  year={2021},
  organization={IEEE}
}

@article{wang2022personalizing,
  title={Personalizing label prediction for GitHub issues},
  author={Wang, Jun and Zhang, Xiaofang and Chen, Lin and Xie, Xiaoyuan},
  journal={Information and Software Technology},
  volume={145},
  pages={106845},
  year={2022},
  publisher={Elsevier}
}

@article{wang2021well,
  title={How well do pre-trained contextual language representations recommend labels for GitHub issues?},
  author={Wang, Jun and Zhang, Xiaofang and Chen, Lin},
  journal={Knowledge-Based Systems},
  volume={232},
  pages={107476},
  year={2021},
  publisher={Elsevier}
}

@dataset{reza2022csiq,
  author       = {Reza, Sayed Mohsin and
                  Mahmud, Saif Uddin and
                  Rahad, Khandoker and
                  Badreddin, Omar},
  title        = {{CSIQ: A Synthesized Dataset of Code Smells, Issues and Quality related Artifacts from Open Source Repositories}},
  month        = aug,
  year         = 2022,
  publisher    = {Mendeley Data},
  version      = {5.0},
  doi          = {10.17632/77p6rzb73n},
  url          = {https://doi.org/10.17632/77p6rzb73n}
}

@inproceedings{romano2021g,
  title={G-Repo: a Tool to Support MSR Studies on GitHub},
  author={Romano, Simone and Caulo, Maria and Buompastore, Matteo and Guerra, Leonardo and Mounsif, Anas and Telesca, Michele and Baldassarre, Maria Teresa and Scanniello, Giuseppe},
  booktitle={2021 IEEE International Conference on Software Analysis, Evolution and Reengineering (SANER)},
  pages={551--555},
  year={2021},
  organization={IEEE}
}

@article{sun2015msr4sm,
  title={MSR4SM: Using topic models to effectively mining software repositories for software maintenance tasks},
  author={Sun, Xiaobing and Li, Bixin and Leung, Hareton and Li, Bin and Li, Yun},
  journal={Information and Software Technology},
  volume={66},
  pages={1--12},
  year={2015},
  publisher={Elsevier}
}

@inproceedings{spadini2018pydriller,
  title={Pydriller: Python framework for mining software repositories},
  author={Spadini, Davide and Aniche, Maur{\'\i}cio and Bacchelli, Alberto},
  booktitle={Proceedings of the 2018 26th ACM Joint Meeting on European Software Engineering Conference and Symposium on the Foundations of Software Engineering},
  pages={908--911},
  year={2018}
}

@article{rees2017better,
  title={Better predictors for issue lifetime},
  author={Rees-Jones, Mitch and Martin, Matthew and Menzies, Tim},
  journal={arXiv preprint arXiv:1702.07735},
  year={2017}
}

@inproceedings{gousios2012ghtorrent,
  title={GHTorrent: GitHub's data from a firehose},
  author={Gousios, Georgios and Spinellis, Diomidis},
  booktitle={2012 9th IEEE Working Conference on Mining Software Repositories (MSR)},
  pages={12--21},
  year={2012},
  organization={IEEE}
}

@inproceedings{gousios2014lean,
  title={Lean GHTorrent: GitHub data on demand},
  author={Gousios, Georgios and Vasilescu, Bogdan and Serebrenik, Alexander and Zaidman, Andy},
  booktitle={Proceedings of the 11th working conference on mining software repositories},
  pages={384--387},
  year={2014}
}

@article{liao2018exploring,
  title={Exploring the characteristics of issue-related behaviors in github using visualization techniques},
  author={Liao, Zhifang and He, Dayu and Chen, Zhijie and Fan, Xiaoping and Zhang, Yan and Liu, Shengzong},
  journal={IEEE Access},
  volume={6},
  pages={24003--24015},
  year={2018},
  publisher={IEEE}
}

@inproceedings{sokol2013metricminer,
  title={MetricMiner: Supporting researchers in mining software repositories},
  author={Sokol, Francisco Zigmund and Aniche, Mauricio Finavaro and Gerosa, Marco Aur{\'e}lio},
  booktitle={2013 IEEE 13th International Working Conference on Source Code Analysis and Manipulation (SCAM)},
  pages={142--146},
  year={2013},
  organization={IEEE}
}

@inproceedings{reza2020modelmine,
  title={Modelmine: a tool to facilitate mining models from open source repositories},
  author={Reza, Sayed Mohsin and Badreddin, Omar and Rahad, Khandoker},
  booktitle={Proceedings of the 23rd ACM/IEEE International Conference on Model Driven Engineering Languages and Systems: Companion Proceedings},
  pages={1--5},
  year={2020}
}

@inproceedings{bissyande2013got,
  title={Got issues? who cares about it? a large scale investigation of issue trackers from github},
  author={Bissyand{\'e}, Tegawend{\'e} F and Lo, David and Jiang, Lingxiao and R{\'e}veillere, Laurent and Klein, Jacques and Le Traon, Yves},
  booktitle={2013 IEEE 24th international symposium on software reliability engineering (ISSRE)},
  pages={188--197},
  year={2013},
  organization={IEEE}
}

@inproceedings{kallis2019ticket,
  title={Ticket tagger: Machine learning driven issue classification},
  author={Kallis, Rafael and Di Sorbo, Andrea and Canfora, Gerardo and Panichella, Sebastiano},
  booktitle={2019 IEEE International Conference on Software Maintenance and Evolution (ICSME)},
  pages={406--409},
  year={2019},
  organization={IEEE}
}

@article{golzadeh2021ground,
  title={A ground-truth dataset and classification model for detecting bots in GitHub issue and PR comments},
  author={Golzadeh, Mehdi and Decan, Alexandre and Legay, Damien and Mens, Tom},
  journal={Journal of Systems and Software},
  volume={175},
  pages={110911},
  year={2021},
  publisher={Elsevier}
}

@inproceedings{bharadwaj2022github,
  title={Github issue classification using bert-style models},
  author={Bharadwaj, Shikhar and Kadam, Tushar},
  booktitle={2022 IEEE/ACM 1st International Workshop on Natural Language-Based Software Engineering (NLBSE)},
  pages={40--43},
  year={2022},
  organization={IEEE}
}

@article{jurado2015sentiment,
  title={Sentiment Analysis in monitoring software development processes: An exploratory case study on GitHub's project issues},
  author={Jurado, Francisco and Rodriguez, Pilar},
  journal={Journal of Systems and Software},
  volume={104},
  pages={82--89},
  year={2015},
  publisher={Elsevier}
}

@inproceedings{yang2017sentiments,
  title={Sentiments analysis in GitHub repositories: An empirical study},
  author={Yang, Bo and Wei, Xinjie and Liu, Chao},
  booktitle={2017 24th Asia-Pacific Software Engineering Conference Workshops (APSECW)},
  pages={84--89},
  year={2017},
  organization={IEEE}
}

@inproceedings{dyer2013boa,
  title={Boa: A language and infrastructure for analyzing ultra-large-scale software repositories},
  author={Dyer, Robert and Nguyen, Hoan Anh and Rajan, Hridesh and Nguyen, Tien N},
  booktitle={2013 35th International Conference on Software Engineering (ICSE)},
  pages={422--431},
  year={2013},
  organization={IEEE}
}

@article{teixeira2019managing,
  title={Managing to release early, often and on time in the OpenStack software ecosystem},
  author={Teixeira, Jos{\'e} Apolin{\'a}rio and Karsten, Helena},
  journal={Journal of Internet Services and Applications},
  volume={10},
  number={1},
  pages={1--22},
  year={2019},
  publisher={Springer}
}

@inproceedings{bertram2010communication,
  title={Communication, collaboration, and bugs: the social nature of issue tracking in small, collocated teams},
  author={Bertram, Dane and Voida, Amy and Greenberg, Saul and Walker, Robert},
  booktitle={Proceedings of the 2010 ACM conference on Computer supported cooperative work},
  pages={291--300},
  year={2010}
}

@techreport{fisher2013utilizing,
  title={Utilizing Atlassian JIRA for large-scale software development management},
  author={Fisher, John and Koning, D and Ludwigsen, AP},
  year={2013},
  institution={Lawrence Livermore National Lab.(LLNL), Livermore, CA (United States)}
}

@article{blischak2016quick,
  title={A quick introduction to version control with Git and GitHub},
  author={Blischak, John D and Davenport, Emily R and Wilson, Greg},
  journal={PLoS computational biology},
  volume={12},
  number={1},
  pages={e1004668},
  year={2016},
  publisher={Public Library of Science}
}

@inproceedings{rios2019methodology,
  title={A methodology for using GitLab for software engineering learning analytics},
  author={R{\'\i}os, Julio C{\'e}sar Cort{\'e}s and Kopec-Harding, Kamilla and Eraslan, Sukru and Page, Christopher and Haines, Robert and Jay, Caroline and Embury, Suzanne M},
  booktitle={2019 IEEE/ACM 12th International Workshop on Cooperative and Human Aspects of Software Engineering (CHASE)},
  pages={3--6},
  year={2019},
  organization={IEEE}
}

@inproceedings{duenas2018perceval,
  title={Perceval: software project data at your will},
  author={Due{\~n}as, Santiago and Cosentino, Valerio and Robles, Gregorio and Gonzalez-Barahona, Jesus M},
  booktitle={Proceedings of the 40th International Conference on Software Engineering: Companion Proceeedings},
  pages={1--4},
  year={2018}
}

\end{document}